\documentclass{aastex631}

\usepackage{CJK}
\usepackage{amsmath}
\usepackage{physics}
\usepackage{fixmath}
\usepackage{upgreek}
\usepackage{textcomp}
\usepackage{amssymb}
\usepackage{upgreek}
\usepackage{tikz}
\usetikzlibrary{arrows}
\usepackage{tabularx}
\usepackage{booktabs,multirow}

\defcitealias{Song2018}{S18}
\newcommand{\BOGOSA}{\citetalias{Song2018}}
\defcitealias{Zhou2019}{Z19}
\newcommand{\BOGOSB}{\citetalias{Zhou2019}}
\defcitealias{Zhou2020}{Z20}
\newcommand{\BOGOSC}{\citetalias{Zhou2020}}

\definecolor{DarkGreen}{rgb}{0.0,0.4,0.0}  
\newcommand{\B}[1]{{\color{blue}{#1}}}
\newcommand{\R}[1]{{\color{red}{#1}}}
\newcommand{\G}[1]{{\color{DarkGreen}{#1}}}
\newcommand{\M}[1]{{\color{magenta}{#1}}}
\newcommand{\C}[1]{{\color{cyan}{#1}}}
\usepackage[normalem]{ulem}
\usepackage{soul}
\usepackage{cancel}

\shorttitle{rotation of magnetic flux rope}
\shortauthors{Zhou et al.}

\received{\today}
\revised{\today}

\begin{document}
\begin{CJK*}{UTF8}{gbsn}
\title{Quantification of the Writhe Number Evolution of Solar Filament Axes}
\correspondingauthor{Zhenjun Zhou}
\email{zhouzhj7@mail.sysu.edu.cn}

\author[0000-0001-7276-3208]{Zhenjun Zhou(周振军)}
\affiliation{Planetary Environmental and Astrobiological Research Laboratory (PEARL), School of Atmospheric Sciences, Sun Yat-sen University, Zhuhai, China}
\affiliation{Key Laboratory of Tropical Atmosphere-Ocean System, Sun Yat-sen University, Ministry of Education, Zhuhai, China}
\affiliation{CAS Key Laboratory of Geospace Environment, University of Science and Technology of China, Hefei, Anhui 230026, China}
\affiliation{CAS Center for Excellence in Comparative Planetology, China}
\author[0000-0002-7018-6862]{Chaowei Jiang}
\affiliation{Institute of Space Science and Applied Technology, Harbin Institute of Technology, Shenzhen 518055, People's Republic of China}
\author[0000-0001-5705-661X]{Hongqiang Song}
\affiliation{Shandong Provincial Key Laboratory of Optical Astronomy and Solar-Terrestrial Environment, and Institute of Space Sciences, Shandong University, Weihai,Shandong 264209, People's Republic of China}
\author[0000-0002-8887-3919]{Yuming Wang}
\affiliation{CAS Key Laboratory of Geospace Environment, University of Science and Technology of China, Hefei, Anhui 230026, China}
\author[0000-0002-0073-7198]{Yongqiang Hao}
\affiliation{Planetary Environmental and Astrobiological Research Laboratory (PEARL), School of Atmospheric Sciences, Sun Yat-sen University, Zhuhai, China}
\author[0000-0002-4721-8184]{Jun Cui}
\affiliation{Planetary Environmental and Astrobiological Research Laboratory (PEARL), School of Atmospheric Sciences, Sun Yat-sen University, Zhuhai, China}
\begin{abstract}
Solar filament eruptions often show complex and dramatic geometric deformation that is highly relevant to the underlying physical mechanism triggering the eruptions. It has been well known that the writhe of filament axes is a key parameter characterizing its global geometric deformation, but a quantitative investigation of the development of writhe during its eruption is still lacking. Here we introduce the Writhe Application Toolkit (WAT) which can be used to characterize accurately the topology of filament axes. This characterization is achieved based on the reconstruction and writhe number computation of three-dimensional paths of the filament axes from dual-perspective observations. We apply this toolkit to four dextral filaments located in the northern hemisphere with a counterclockwise (CCW) rotation during their eruptions. Initially, all these filaments possess a small writhe number ($\leqslant 0.20$) indicating a weak helical deformation of the axes. As the CCW rotation kicks in, their writhe numbers begin to decrease and reach large negative values. Combined with the extended C\u{a}lug\u{a}reanu theorem, the absolute value of twist is deduced to decrease during the rotation. Such a quantitative analysis strongly indicates a consequence of the conversion of twist into writhe for the studied events.

\end{abstract}

\keywords{Sun: filament --- Sun: corona --- Sun: prominences}
\emph{Online-only material}: animations, color figures
\section{Introduction}
Solar magnetic flux ropes (MFRs), defined as a group of twisted magnetic field lines wrapping around a common axis,  are core structures driving solar eruptions such as flares and coronal mass ejections.
The writhe of the MFR's axis is a fundamental parameter for characterizing its global geometric deformation, which has been found highly correlated with the dynamic property of MFRs. 
Many observational features in the corona reveal MFRs with writhed axes from different observational wavelengths and perspectives.
For example, the presence of S-shaped structures on the solar disk, like sigmoids observed in the extreme ultraviolet (EUV) and soft X-ray (SXR) passbands, 
are expected to be the manifestation of enhanced current density in the MFRs \citep{Kliem2004,Gibson2006}.
The filament, often observed underneath a cospatial sigmoid, is embedded in the bottom of the MFR \citep{Cheng2014}. Thus the shape of a filament is an alternative substitution to trace the MFR's axis \citep{Zhou2017}.

Besides a distorted axis, filament often exhibits an internal twisted structure. By tracing the threads of filament about its axis in high-resolution observations, like the Swedish 1-m Solar Telescope \citep[SST,][]{Scharmer2003} and the
New Vacuum Solar Telescope \citep[NVST,][]{Liu2014}, spiral structures and motions are frequently observed \citep[e.g.,][]{Schmieder2000}. 
The bright and dark signals as observed on opposite sides of a filament in both the H$\alpha$ and its corresponding line-of-sight velocity (Doppler) observations suggest that interwinding threads follow the helical field lines inside filament.
These observational facts exhibit helical threads wrapping around the axis, in favor of the existence of considerable twist, and an overall distorted shape, indicating a writhed axis.

When erupted, 
the spinning motions observed in jets, as well as apex rotations in erupting filaments, suggest 
a relaxation of magnetic twist, thus 
lowering the magnetic energy of the MFR by reducing the bending of its field lines.
In particular, under the ideal MHD constraint of helicity conservation, the rotation of the filament axis in the course of its rise is generally interpreted as a consequence of the conversion of twist into writhe.
Numerical simulations \citep{Kliem2012,Hassanin2016} recreate this rotation motion in a kink-unstable magnetic flux rope. A kink instability (KI) serves as the key mechanism leading to the rotation. It refers to the helical instability of the MFR which commences when the twist of the rope exceeds some critical value \citep{Torok2004}. This instability has received great attention due to its good quantitative agreement with many well-observed events \citep{Torok2005,Williams2005,Kliem2012}.  Another rotation mechanism, the Lorentz force due to the external sheared field, is also addressed as a major contributor to driving the filament rotating \citep{Isenberg2007,Kliem2012}. 

The observation shows that 
clockwise (CW)/counterclockwise (CCW) rotation is associated with sinistral/dextral filaments \citep{Green2007,Zhou2020}, no matter whether the rotation is triggered by the kink instability or the external sheared field \citep{Isenberg2007,Kliem2012}. However, this leaves a mystery\citep{Zhou2022}: 
sinistral/dextral filaments often exhibit a forward/reverse S shape, 
while through a CW/CCW rotation, the filament spine is straightened and even over-rotated to reverse its initial shape.
During this process, the writhe number of the spine seems to experience a decreasing process, though such an variation has never been systematically investigated.

Revealing the variation of the writhe number for a filament axis during its rotation is a key step in understanding which mechanism triggers the eruptions. However, 
this temporal evolution has hardly been quantified, even for numerical simulations that provide the three-dimensional (3D) magnetic field of erupting, kinked flux ropes \citep{Linton1998}. 
The general definition of writhe number makes its computation a relatively complicated task. Moreover, in reality, it is difficult to obtain the 3D geometry
of a filament axis to limited observations.

In this Letter, we report the development of a Writhe Application Toolkit (WAT) for computing precisely the writhe number of solar filaments and its application to quantitative investigation of
writhe number variation in four filament eruptions.
These eruptions are recorded from dual-perspective observations with the dramatic development of writhe during rotations.  
In the sections that follow, the introductions of the instruments and methods are presented in \S \ref{sec:Obs}; 
A detailed analysis of a typical example is given in \S \ref{sec:Cal}.  We finally summarize and discuss our results in \S \ref{sec:res} and \S \ref{sec:dis}.

\section{Observation and Analysis} \label{sec:Obs}
\subsection{Instruments and Methods}
Our computation is based on the dual-perspective observations from two spacecraft. The Solar TErrestrial RElations Observatory \citep[STEREO,][]{Kaiser2008} mission consists of two spacecraft placed in heliocentric orbits ahead (STEREO-A) and behind (STEREO-B) of Earth. As viewed from the sun, these two spacecraft separated at approximately $22^\circ$ per year with Earth in opposite directions. Meanwhile the Solar Dynamics Observatory \citep[SDO,][]{Pesnell2012} was placed in the geosynchronous orbit. 
The paired Extreme Ultraviolet Imager \citep[EUVI;][]{Wuelser2004} telescopes onboard STEREO can observe a filament in narrow extreme-UV (EUV) passbands including 195 {\AA} (formation temperature, $T_f \simeq 1.6 \times 10^6 K$)  and 304 {\AA} ($\simeq 6 \sim 8 \times 10^5 K$). The typical cadences for these two filters are 5 and 10 minutes. 
The Atmospheric Imaging Assembly \citep[AIA;][]{Lemen2012} on board SDO has similar filters as 193 {\AA} ($\simeq 1.58 \times 10^6 K$) and 304 {\AA} ($\simeq 10^5 K$) channels with a higher cadence (12s).
Their simultaneous observations in a multi-view setting allow us to perform stereoscopic triangulation on a solar filament.
Meanwhile the Helioseismic and Magnetic Imager \citep[HMI;][]{Hoeksema2014} onboard SDO provides the vector magnetograms which is conducive to the determination of filament chirality.

To get the 3D configuration of a filament, we employ the scc\_measure widget application \citep{Thompson2009} in the Solar SoftWare, which uses tie-pointing method for reconstruction \citep{Inhester2006}. 
We first select a point along the filament spine in, for example, the AIA image,
then this routine will calculate a path representing the line of sight from another perspective displayed in, for example,  the EUVI image. 
According to the dynamic evolution and emission characteristics,
the same feature point along this path can be identified manually.
Its 3D coordinates (heliographic longitude, latitude, and radial distance in solar radii) are then calculated.
The steps above are repeated until the entire filament axis including its footpoints is reconstructed.  

The positional data of the filament axis are then smoothed and interpolated using 3D cubic B-splines, with the aid of the SciPy {\tt\string interpolate.splprep} module and the smoothing factor is set to $0.0005$. The position of the spline is evaluated using the {\tt\string interpolate.splev} module.

The reconstruction, smoothing, and interpolating of the filament axis at all time-frames follow the same processing procedures.
Thus, during the eruption, the evolution of the filament axis, including the rotation and expanding motion, could be studied in three dimensions.

A new definition of writhe, termed the polar writhe $\mathcal{W}_p$ \citep{Berger2006}, is designed to measure the writhe of open curves with endpoints on a boundary plane. This definition is especially appropriate to quantify the helical deformation of the filament axis. 
The computation of the polar writhe $\mathcal{W}_p$ is decomposed into local and nonlocal components:
\begin{equation} 
\mathcal{W}_{p}(\textbf{r})=\mathcal{W}_{pl}(\textbf{r})+\mathcal{W}_{pnl}(\textbf{r}), \label{eqwr0}
\end{equation} 
Here the curve is divided into $n+1$ pieces at $n$ turning points ($ds/dz=0$) in vertical direction \textbf{\textit{z}}. 
The local writhe ($\mathcal{W}_{pl}$) measures helical coiling in each individual piece,
where the nonlocal polar writhe ($\mathcal{W}_{pnl}$) gives a quantitative calculation of the global geometric relations between these pieces. $\mathcal{W}_{pl}$ and $\mathcal{W}_{pnl}$ can be calculated as \citep{Berger2006}:

\begin{equation} 
\mathcal{W}_{pl}=\mathop{\sum_{i=1}^{n+1}} \frac{1}{2\pi}\int_{z^{min}_{i}}^{z^{max}_{i}}\frac{\boldsymbol{z} \cdot \boldsymbol{T}_i\times \frac{d\boldsymbol{T}_i}{dz}}{1+|\boldsymbol{z}\cdot \boldsymbol{T}_i|}dz, \label{eqwr1}
\end{equation}

\begin{equation} 
\mathcal{W}_{pnl}=\mathop{\sum_{i=1}^{n+1}\sum_{j=1}^{n+1}}_{i\neq j} \frac{\sigma_{ij}}{2\pi}\int_{z^{min}_{ij}}^{z^{max}_{ij}}\frac{d\Theta_{ij}}{dz}dz, \label{eqwr2}
\end{equation} 

where $i$,$j$ are two different pieces, 
$\mathbf{T}_i=d\mathbf{x}_i/ds$ is the unit tangent vector of the piece $\mathbf{x}_i$,
the indicator function $\sigma_{ij} =+1$ if these two pieces are moving towards the same direction, $\sigma_{ij} =-1$ if two pieces are moving in opposite directions. 
$\Theta_{ij}$ is the orientation of the relative position vector ($\mathbf{r}_{ij}=\mathbf{x}_j-\mathbf{x}_i$) with respect to the $x$ axis.
Based on Equations (\ref{eqwr0})-(\ref{eqwr2}), a tool for the numerical computation of the polar writhe is available online \citep{Prior2016}\footnote{\url{https://www.maths.dur.ac.uk/~ktch24/code.html}}. We applied this code to the calculation of the  writhe for the filament axis.  

All these processes above are integrated into a publicly available package called Writhe Application Toolkit\footnote{\url{http://sysu-pearl.cn:8080/writhecode}} (\href{http://sysu-pearl.cn:8080/writhecode}{WAT}).

\subsection{Selection of Events}
Through browsing image data for solar filament eruption in the past 12 years since the launch of SDO,
four cases are picked out following the criteria below:
(1) An obvious rotation motion can be observed during the filament eruption.
(2) The entire filament including its footpoints is required to be observed from at least two different perspectives. 
The SDO/AIA observation from Earth perspective is a necessity as its high cadence ensures simultaneously imaging the Sun with STEREO.
The selected filament eruptions and their properties are listed in Table~\ref{tab:mytable}, 
included the direct observational characteristics and inferred properties.
Details of the selected events (including the dual-perspective imaging observations and the reconstructed position of the filament axes) are available at this \href{http://sysu-pearl.cn:8080/solarEruptive}{catalog website}\footnote{\url{http://sysu-pearl.cn:8080/solarEruptive}}.

\section{Calculation of Writhe Evolution}\label{sec:Cal}

\begin{figure*} 
      \vspace{-0.03\textwidth}    
      \centerline{\hspace*{0.00\textwidth}
      \includegraphics[width=1.0\textwidth,clip=]{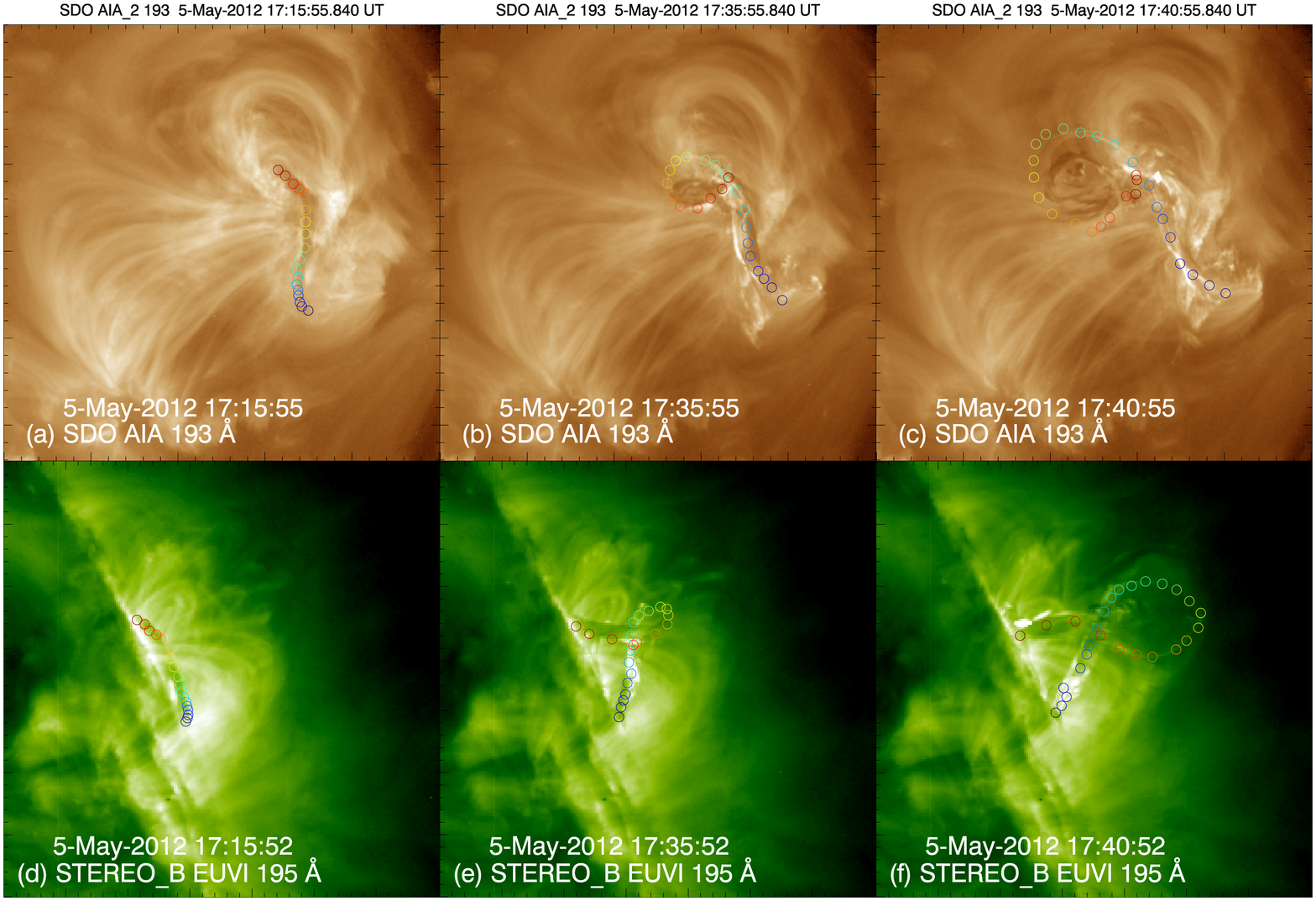}
      }
\caption{
Filament rotation during its eruption on 2012 May 5. 
(a)-(c) The AIA 193 {\AA} images at 17:15 UT, 17:35 UT, and 17:40 UT displaying the top view of the evolution of this filament rotation. (d)-(f) The EUVI B 195 {\AA} images showing the side view of the filament rotation.
The color-coded points mark identical points on the sun as seen from STEREO-B and SDO, respectively. An animation of AIA 193 {\AA} and EUVI 195 {\AA} direct images is available online to show the rotation of the filament, the video covers the time from 17:15:07 UT to 17:50:55 UT with a cadence of 12s, and} its duration is 9 s.
 \label{f1}
\end{figure*}

\begin{figure*} 
      \vspace{-0.03\textwidth}    
      \centerline{\hspace*{0.00\textwidth}
      \includegraphics[width=1.0\textwidth,clip=]{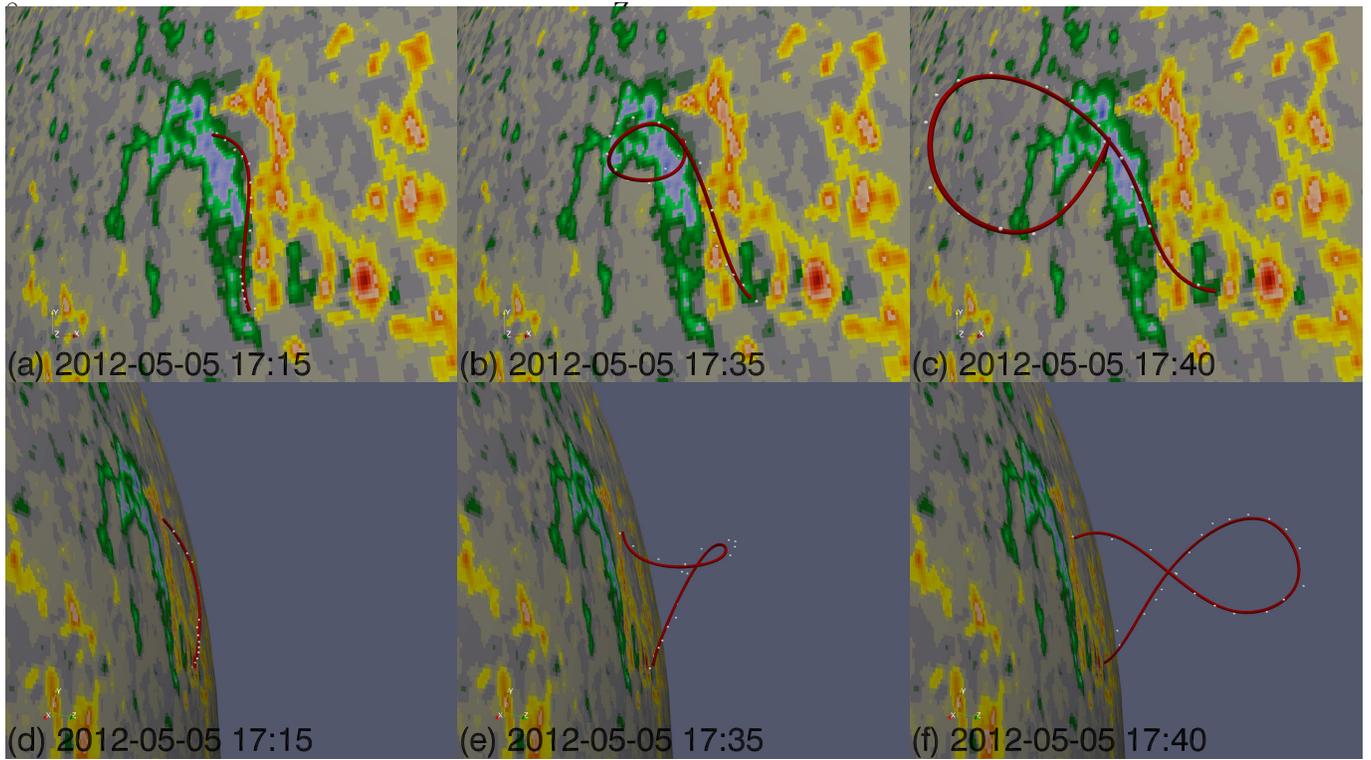}
      }
\caption{
The 3D reconstruction of the filament axis during the eruption viewed from the top (a)-(c) and side (d)-(f). White solid points are measured by triangulation from dual viewing angles, the red solid curves are smoothed and interpolated from white points,
and the bottom boundaries are the projected HMI daily update radial synoptic map.
} \label{f2}
\end{figure*}

Among these four cases in Table~\ref{tab:mytable}, we take Case~\#3 as an example (see Figure~\ref{f1}). 
This filament has dextral chirality as inferred from its left-skewed drainage sites (Figure~\ref{f2}(a)) \citep{Chen2014,Zhou2020}.
The filament eruption occurred on 2012 May 5 at 17:15 UT. It is observed in NOAA Active Region (AR) 11474 at N16E35 from SDO view. 
It is also captured by STEREO-B from the limb view. At that time, the separation angle between SDO and STEREO-B is $118^{\circ}$. 
In the SDO AIA 193 {\AA} channel (Figure~\ref{f1}(a)), the initial shape of the filament takes on a reverse S shape. When erupting, the filament shows an obvious CCW rotation about the rising direction (Figure~\ref{f1}(a)-(c)), transforming the morphology of filament from a $\Uplambda$ shape to a reversed-Y shape as seen in side view (STEREO-B 195 {\AA}, Figure~\ref{f1}(d)-(f)).

The axis of the filament is identified from SDO and STEREO-B images for each snapshot (the color coded points in Figure~\ref{f1}).
Then the 3D positional information of the axis can be determined by the triangulation. Applying 3D cubic B-splines mentioned above, a smoothed and interpolated filament axis is obtained.  Repeating the procedure frame by frame, the evolution of the filament axis during eruption can then be acquired, as illustrated by the red spine in Figure~\ref{f2}.

For each time frame, the writhe of the axis curve is computed using the Prior \& Neukirch code. 
We applied the same processing procedures to the other 3 cases in Table~\ref{tab:mytable}, and
the writhe evolutions of these filament axes during the rotation are demonstrated in Figure~\ref{f3}. 
It should be noted that Case~\#2 and \#3 are reconstructed with AIA 193 {\AA} and EUVI\_B 195 {\AA} images since they have a higher effective cadence of 5 minutes.
Limited by the indistinctness in 195 {\AA} or 193 {\AA} images, the reconstruction of filaments for Case~\#1 and \#4 has to consult 304 {\AA} images (10 minutes cadence).
We repeat the reconstruction 5 times for each case to reduce and estimate the errors introduced by measurement.  

\section{Results} \label{sec:res}
\begin{figure*} 
     \vspace{-0.0\textwidth}    
     \centerline{\hspace*{0.00\textwidth}
     \includegraphics[width=1.0\textwidth,clip=]{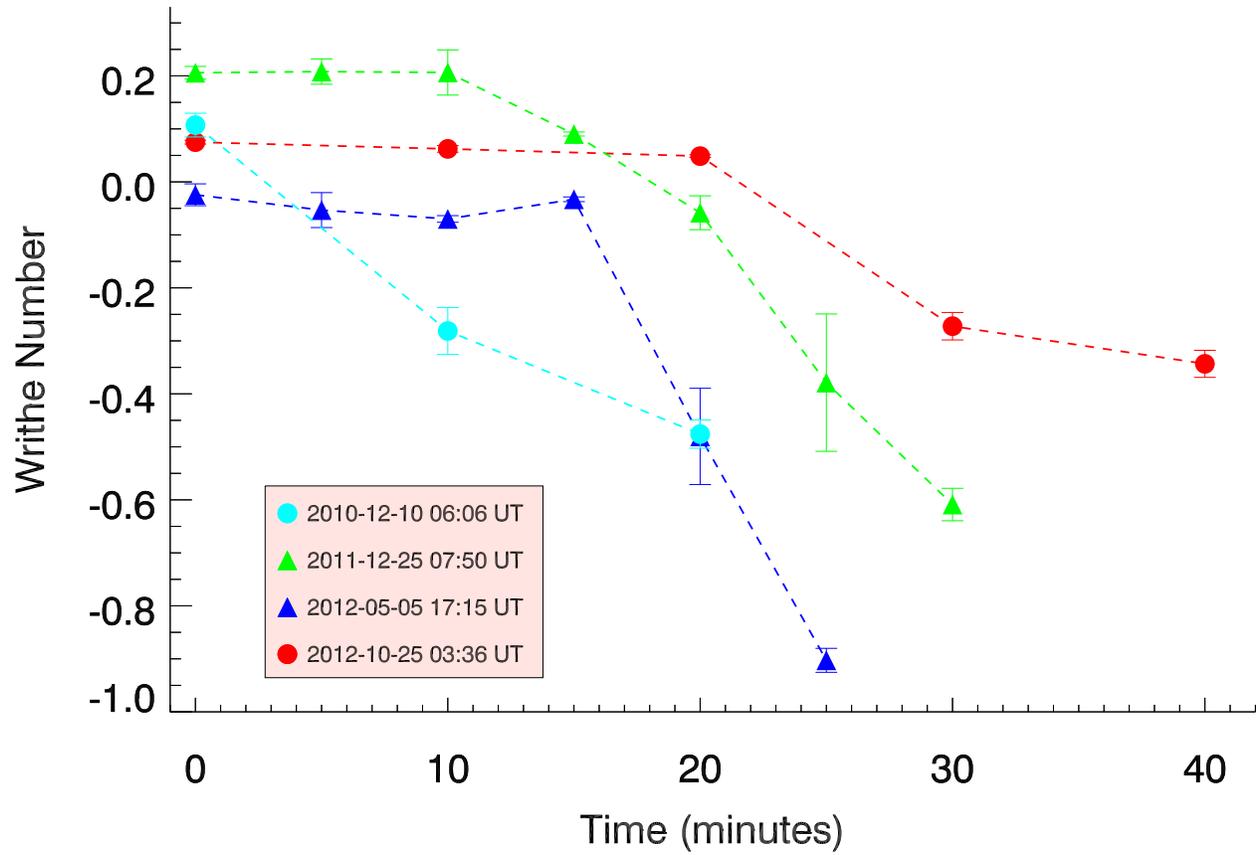}
               }
\caption{
The variation of writhe number for filament axes during their eruption. Each event is marked by a different color. Events labeled with triangle symbols are reconstructed from EUVI 195 {\AA} and AIA 193 {\AA} images and with circle symbols reconstructed from EUVI and AIA 304 {\AA} images. The error of the writhe number (marked by the vertical error bar) mainly comes from the uncertainty of feature points location, which is taken as the standard deviation of 5 measurements.
} \label{f3}
\end{figure*}

\begin{figure*} 
      \vspace{-0.03\textwidth}    
      \centerline{\hspace*{0.00\textwidth}
      \includegraphics[width=0.6\textwidth,clip=]{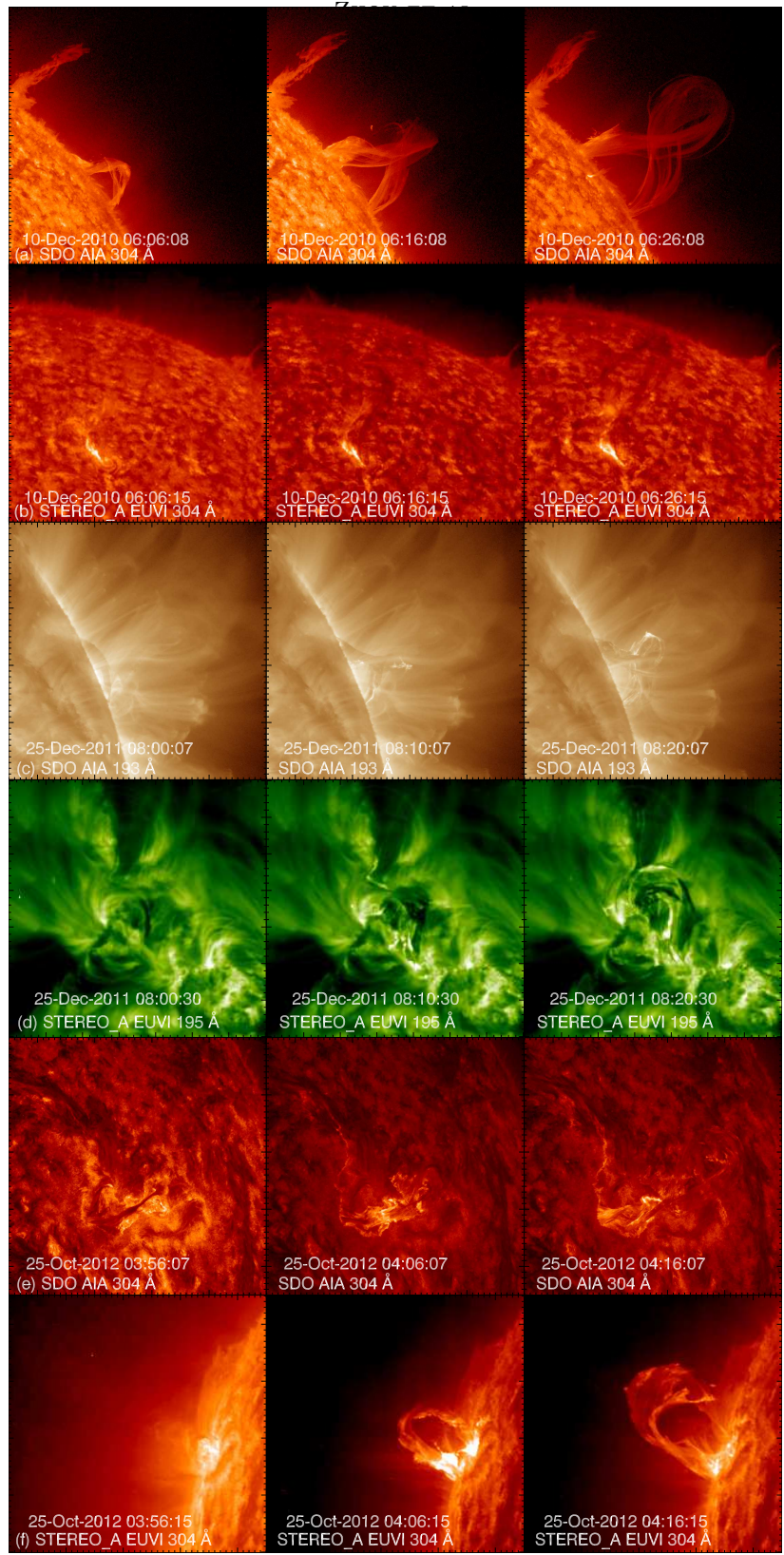}
      }
\caption{
Filament rotation during its eruption for the other 3 cases from dual-perspective observations. Panels (a) and (b) correspond to cyan dashed line in Figure~\ref{f3}, panels (c) and (d) correspond to green dashed line, and panels (e) and (f) correspond to red dashed line.
} \label{f4}
\end{figure*}

The variations of writhe and their errors during eruptions are displayed in Figure~\ref{f3}. 
Initially, all these filaments possess a small writhe number ($\leqslant 0.20$) indicating a weak helical deformation of the axis. 
And the writhe trajectory is relatively shallow, the filament is slow evolving without large changing in geometry. The very steep drop in writhe in Figure~\ref{f3} for each case means a strong kinking for the filament geometry, this can be verified in Figure~\ref{f4}: the filaments in middle subfigures are experiencing transformed into an inverse-$\gamma$ shape from a side view. When the filaments have sufficiently risen,  the nonlocal writhe dominates.  
Actually, this nonlocal writhe measures the apex rotation angle (a nonlocal writhe of $\pm2$ indicates one full turn of $\mp360^\circ$ for a filament apex rotation)\citep{Torok2010}.
Meanwhile, Figure~\ref{f3} shows an interesting fact that the writhe evolutions of the triangle plots (cases~\#2 and \#3) are very similar to each other but different form the circle ones (cases~\#1 and \#4). We consider that this results from sample selection bias for the cadence of EUVI 304 {\AA} is 10 minutes and that of EUVI 195 {\AA} is 5 minutes. This implies, to get at least 3 frames, a rotation we selected in EUVI 304 {\AA} must last for at least 20 minutes, but for cases selected in EUVI 195 {\AA}, duration can be just 10 minutes. Therefore, the writhe evolutions selected in EUVI 195 {\AA} (cases~\#2 and \#3) seems steeper than those in EUVI 304 {\AA} (cases~\#1 and \#4).
Noting that these four cases are located in the northern hemisphere, 
and due to the mirror symmetry for observational properties between the northern and southern hemispheres,
the initial shape and the sense of rotation exhibit opposite behaviors in the southern hemisphere \citep{Zhou2020}. Thus, 
in the souththern hemisphere the writhe should increase during filament rotation.

\section{Discussion} \label{sec:dis}
In this work, we provide a set of processes to reconstruct, smooth, and interpolate the axis of a filament. After that,
aided by the Prior \& Neukirch code, the writhe of the axis curve is computed.
This suite of processes is assembled as the Writhe Application Toolkit (WAT). 
We apply this toolkit to cope with four filament eruption cases with an obvious rotation motion, and the evolvement of writhe number for solar filament axes is achieved.

These four cases are identified as dextral filaments \citep{Song2018,Zhou2019,Zhou2020}, implying a negative magnetic helicity contained \citep{Chae2000}.
Helicity can be decomposed into writhe and twist contributions \citep{Berger1984a}, thus for a negative helicity filed with little or positive writhe ($\mathcal{W}_{p}$ ), the sign of the twist ($\mathcal{T}_{w}$) must be negative. 
\citet{Berger2006} extend  C\u{a}lug\u{a}reanu theorem to general open curves that are anchored on a plane, like magnetic flux tubes in the solar corona:
the net-winding ($\mathcal{L}_{p}=\mathcal{W}_{p} + \mathcal{T}_{w}$) is invariant to all motions (as long as the two curves do not intersect and a single curve does not reconnect), so a decrease in writhe must be acompensated by an increase in twist \citep{Berger1984b,Linton1998}.

Combining the writhe evolution of the filament axis, the evolution of the twist evolution for the studied events can be divided into three stages as follow: 
(1) Initially, the twist number is negative, indicating that the magnetic field lines are left-hand wrapping around the filament axis; 
(2) As a CCW rotation kicks in, $\mathcal{W}_{p}$ decreases to $0$. Based on the extended  C\u{a}lug\u{a}reanu theorem, there is a corresponding increase in $\mathcal{T}_{w}$. Considering the negative sign, the absolute value of $\mathcal{T}_{w}$ should decrease during the rotation.
(3) As the CCW rotation continues, $\mathcal{W}_{p}$ decrease from 0 to a large negative value,
and the variation trend of $\mathcal{T}_{w}$ is consistent with that in stage 2. It is worth noting that the potential input through or at the photosphere is not considered,
due to the small rotation ($\leqslant 25^{\circ}$, corresponding to 0.069 changes in the net-winding $\mathcal{L}_{p}$  for the sample case~\#3) derived from the foot points shifting of the filament. Also the rotation of the filament around the foot points at the photosphere is too small to identified.

This quantitative analysis of MFR's $\mathcal{W}_{p}$ and $\mathcal{T}_{w}$ provides an alternative answer to a long-lasting mystery:
Generally, sinistral/dextral filaments (located in the southern/northern hemisphere) exhibit a forward/reverse S shape, and when erupted, they often rotate clockwise/counterclockwise when viewed from above, and thus the overall shape of the filaments seem to be straightened or even reversed.
In our analysis, the process of being straightened for a reverse S shape filament corresponds to the decreasing of a initially positive writhe to $0$. 
This process is not against the conversion of twist into writhe in the course of flux rope instabilities. The absolute value of $\mathcal{T}_{w}$ constantly decreases during the rotation.

In these four cases, the writhe and projected S shape is not unique, and the configurations of the filament axes, including the skewness of filament legs, the filament middle section distorted by the polarity inversion line (PIL), the relative height, the position of the highest point, and the position of the crossing point,  all contribute to the writhe number \citep{Torok2010,Xu2020,Zhou2020}. 
In summary, 4 filament eruptions in the northern hemisphere with obvious rotation motion are studied in the dual-perspective observations from STEREO and SDO. The evolutions of filament axes are obtained  with the help of the 3D reconstruction and 3D cubic B-splines. The polar writhe ($\mathcal{W}_{p}$) is employed to quantitate the contortion of the filament axis during the rotation. During the eruption, $\mathcal{W}_{p}$ decreases from a small value ( positive or near $0$) to a large negative value. Correspondingly, based on the extended  C\u{a}lug\u{a}reanu theorem,  the twist ($\mathcal{T}_{w}$) of a dextral filament (negative helicity contained) should increase.  Considering the negative sign of twist, its absolute value decrease during the rotation. This is consistent 
with the transformation of twist into writhe in a kink-unstable MFR.

\section{acknowledgements} \label{sec:ack}
The authors appreciate discussions with Rui Liu, Xin Cheng, and Quanhao Zhang.
We acknowledge the \emph{SECCHI}, \emph{AIA}, and \emph{HMI} consortia for providing  excellent observations.

This work is supported by the B-type Strategic Priority Program XDB41000000 funded by the Chinese
Academy of Sciences. The authors also acknowledge support from the National Natural Science Foundation of China (NSFC 42004142,42274203), Open Research Program of CAS Key
Laboratory of Geospace Environment, Science and Technology Project 202102021019 in Guangzhou, the Fundamental Research Funds for the Central Universities (grant No.HIT.BRETIV.201901), and Shenzhen Technology Project JCYJ20190806142609035.


  \begin{deluxetable}{ccccccccc}
   \tablenum{1}
   \tablecaption{Properties of the Selected Filaments}
   \label{tab:mytable}
   \tablewidth{0pt}
   \tablehead{
   \colhead{No.} & \colhead{Start Time} & 
   \multicolumn{3}{c}{Observations} & \multicolumn{3}{c}{Inferred Properties}  & \colhead{Ref.$^\dag$} \\
   \cmidrule(lr){3-5} \cmidrule(lr){6-8}
   \colhead{} & \colhead{} & \colhead{Hemisphere} & \colhead{Projection Shape} & \colhead{Rotation} & \colhead{Chirality} & \colhead{Initial Writhe} & \colhead{Trend} & \colhead{} 
   }
   \decimals
   \startdata
1 & 2010-12-10 06:06 & N & Z & CCW & Dextral &  0.107   & Decrease   & \BOGOSA{} \\ 
2 & 2011-12-25 07:50 & N & Z & CCW & Dextral &  0.206   & Decrease   & \BOGOSC{} \\
3 & 2012-05-05 17:15 & N & Z & CCW & Dextral & -0.024   & Decrease   & \BOGOSB{},\BOGOSC{} \\
4 & 2012-10-25 03:36 & N & S & CCW & Dextral &  0.075   & Decrease   & \BOGOSB{},\BOGOSC{} \\ [1ex]
   \enddata
   $^\dag$ Previous investigations of the filament eruption. \BOGOSA{} refers to \citet{Song2018}, \BOGOSB{} to \citet{Zhou2019}, and \BOGOSC{} to \citet{Zhou2020}.
   \end{deluxetable}

\end{CJK*}
\end{document}